\documentclass{article}    

\title{ \bf  Expected Shortfall \\ \bf as a Tool for Financial Risk Management}  
\author{\small \it Carlo Acerbi\footnote{\tt cacerbi@abaxbank.com, +39 02 77426 402}, Claudio Nordio\footnote{\tt cnordio@abaxbank.com, +39 02 77426 407} and Carlo Sirtori\footnote{\tt csirtori@abaxbank.com, +39 02 77426 401} \\  
 \small\it 
Abaxbank, Corso Monforte 34,  20122 Milano Italy}

\begin{document}           
\maketitle                 


\centerline{\bf \large Abstract}

\begin{quotation}
\em \small \noindent We study the properties of Expected Shortfall from the point of view of financial risk management. This measure --- which emerges as a natural remedy in some cases where VaR is not able to distinguish portfolios which bear different levels of risk --- is indeed shown to have much better properties than VaR. We show in fact that unlike VaR this variable is in general subadditive and therefore it is a Coherent Measure of Risk in the sense of reference \cite{artzner}
\end{quotation}

\vspace{15mm}

In this paper we  review some classical arguments which arose in the last years in the debate on Value at Risk (VaR) as a  measure for assessing the financial risks of a portfolio and we  analyze an alternative measure of risk, which is a version of the Expected Shortfall used in  Extreme Value Theory.
In the second part of the paper the comparison between the two risk measures will be made on a more technical ground by analyzing some mathematical properties that play a crucial role in the definition of a risk measure.

We begin with a paradox assuming that the reader is familiar with the concept of VaR.

\section{A paradox} \label{paradox}

Consider a portfolio A (made for instance of long option positions) of value 1000 Euro with a maximum downside level of 100 Euro and suppose that the worst 5\% cases on a fixed time horizon T are all of maximum downside. VaR at 5\% on this time horizon  would then be 100 Euro.
Consider now another portfolio B again of 1000 Euro which on the other hand invests also in strong short futures positions that allow for a potential unbounded  maximum loss. We could easily choose B in such a way that its VaR is still 100 Euro on  the time horizon T.

However:
\begin{itemize}
	\item In portfolio A the 5\% worst case losses are all of 100 Euro.
	\item In portfolio B the 5\% worst case losses range from 100 Euro to some arbitrarily high value.
\end{itemize}
{\bf Which portfolio is more risky ? According to VaR 5\% they bear the same risk !} 

We'll come back on this example later. We first review some classical arguments of the debate around VaR.

\section{The classical ``Pros and Cons'' of Value at Risk}  \label{prosandcons}

In financial risk management VaR has certainly represented a significative step forward with respect to more traditional  measures mostly based on sensitivities to market variables (the ``Greeks'').
The strength of VaR relies in
\begin{enumerate}
\item {\em VaR applies  to any  financial instrument and it is expressed in the same unit of measure}, namely in ``lost money''. Greeks on the contrary are  measures created {\em ad hoc} for specific instruments or risk variables and are expressed in different units.
The comparison of relative riskiness between, say, an equity portfolio and a forex portfolio is not easy with Greeks, while it is a straight comparison knowing their VaR's.
\item {\em VaR includes an estimate of future events} and allows one to convert in a single number the risk of a portfolio. Greeks on the contrary essentially amount to ``what if'' variables. Saying for instance that one loses 1 Euro if interest rates raise of 1 bps, one still wonders ``how likely it is'' that interest rates do indeed raise of 1 bps. VaR on the contrary does exactly this job.
\end{enumerate}
The difficulties encountered when computing VaR on financial portfolios come exactly from the above mentioned ambitious tasks:
\begin{enumerate}
	\item  including in the analysis the whole set of risk variables which affect a portfolio.
	\item  estimating correctly the probabilities of future market events 
\end{enumerate}
In real cases, due to the complexity of computational aspects and to the delicacy of the estimate of market probabilities, in order to compute VaR one has to resort to (sometimes strong) assumptions both on the functional dependence of financial instruments from risk drivers (first--second order approximations \ldots) and on the probability distributions (historical VaR, parametric VaR \ldots).

We mention the fact that sometimes it's useful to decouple the risks associated to different risk drivers. VaR can be then computed ``switching on'' just some class of risk drivers, holding all the remaining fixed. One then speaks in this case of partial VaR's like  ``Interest Rate VaR'' (IRVaR), ``Forex VaR'' (FXVaR), ``Equity VaR'' (EQVaR), ``Credit VaR'' (CVaR) and so on.

In the case of complex portfolios exposed to many risk variables such as in financial institutions, the computation of VaR can often be a formidable task. A challenging aspect is due to the fact that the computation can not be split into separate sub--computations due to the two--fold non--additivity of VaR:
\begin{enumerate}
	\item{Non--additivity by position:} given a portfolio made of two subportfolios, total VaR is not given by the sum of the two partial VaR's, with the consequence that adding a new instrument to a portfolio often make it necessary to recompute the VaR for the whole portfolio.
	\item{Non--additivity by risk variable:} For a portfolio depending on multiple risk variables, VaR is not the sum of partial VaR's. So, for instance, for a convertible bond, VaR is not simply the sum of its IRVaR and EQVaR.
\end{enumerate}
In both cases, in the case of normal distributed returns of a portfolio, one can  show that the ``non--additivity''
is actually a ``sub--additivity'': total VaR is always less or equal than the sum of partial (by position or by risk driver) VaR's. On the possibility of extending the sub--additivity of VaR beyond the Gaussian case we will say more in the sections below.

Even if non--additivity raises serious computational difficulties, it is however the direct sign of one of the most interesting aspects of VaR as an instrument for risk analysis, namely its ability to exhibit better than any Greeks the advantages due to {\em diversification} of financial instruments and risk drivers. VaR is in fact sensitive to the  hedging  effect of different positions and the mutual correlation effect of risk drivers. The sub--additivity of VaR in a Gaussian world embodies the common belief that {\em diversifying lower risks}.

We are now in a position to understand better the most frequent criticisms which are used against VaR, which in the past gave rise to harsh battles between opposite factions (see for instance \cite{Jorion,Taleb}).
\begin{itemize}
\item{\em 
``VaR always come late when the damage is already done'':} this well known adage comes from the fact that in order to estimate market probabilities it's a common habit to calibrate future scenarios on past market data. To make an example, it's clear that the day before some market turbulence, the parameter estimates will not be able to forecast the sudden jump of volatilities and VaR will inevitably underestimate the risks, noticing the increase of risks only some days later, when the storm is over.

It must be said however that this objection should not be addressed to VaR itself, but rather to a certain (yet widespread) way of computing VaR by estimating future probabilities from the past.
In a trading desk for example this could be avoided by requiring proper volatility inputs from the trader themselves who can timely foresee future bumps of volatility from the news or from market fundamentals. 
\item{\em ``VaR makes no sense'':} it may appear defeatism, but VaR estimates, especially in the case of complex portfolios may be such a difficult computational task that  the final number may have no statistical value at all. It's not difficult 
in these cases to see absurd VaR figures which have nothing to do with the risks of the portfolio.
Among the factors which can make the estimate particularly difficult we can identify:
\begin{enumerate}
	\item the complexity of financial instruments
	\item the dimension of portfolio
	\item the assessment of market probabilities
	\item the approximations introduced to speed up the computation
	\item the statistical error on the estimate of VaR. VaR in facts, as most quantile--based variables is affected by large statistical error, in particular when the chosen probability level is very small. \cite{Embrechts}.
\end{enumerate}
The set of assumptions, approximations and statistical errors introduced in the computation can be so large that the final result is nonsense due to  its wide confidence level. In other words, behind a reassuring 
{\em ``VaR= $100$ Euro''} an actual reckless {\em ``VaR= $100 \pm 10000$ Euro''} can be hidden.

These observations can seem pessimist. But as for any other statistical estimate the correct thing to do is very clear: one should not trust in it unless all the approximations and assumptions made for the estimate are under control and unless the user of the final result is aware of the hypotheses which have been made.

The worst thing to do also in this case is unfortunately the most widespread: producing VaR estimates with no mention of the hypotheses underlying and pretending with some unaware user that the numbers are reliable.
\end{itemize}
From what we said it's clear that VaR is a rather sophisticated and complex tool to be used. Using VaR requires not only good mathematical skills, but also much care in the interpretation of results and transparency in declaring the scope of its validity. As any sophisticated analysis tool, VaR can provide precious information when used correctly, but can be very dangerous in the wrong hands.

\section{Why VaR ?} \label{why}

It may seem surprising, but all we said above speaking of VaR could be said with no modification for a large class of statistical variables, namely the whole class of statistics of the left tail of the portfolio returns distribution. For all of these variables, the computational effort needed to model the probability distribution and to create a sample of events is very much the same. Once this is made, any statistical variable of the left tail can be determined with no additional effort\footnote{We neglect at this level the important issues related to the error of the estimate.}. One therefore is led to wonder why the financial market has chosen VaR in this wide class and not some other statistics. 

Is  really VaR the most representative variable for characterizing the risks associated with extreme events of a portfolio ? Why VaR has been chosen worldwide by financial risk managers and bank regulators as the proper measure for controlling the risks of financial portfolios ?

Let's begin with some elementary considerations on the very definition of VaR. As any practitioner knows, in order to compute VaR one has to choose a specified level of confidence  selecting the set of ``worst cases'' under consideration and a time horizon over which the estimates of future Profit \& Loss are made. With no loss of generality we will consider in what follows a probability of 5\% and a time horizon of 7 days.

A definition that is often used for VaR is the following.
\begin{quotation}
\em 
``VaR is the maximum potential loss that a portfolio can suffer in the 5\% worst cases in 7 days''
\end{quotation}
Of course the above definition is wrong. The following correct version is however seldom used because it sounds odd and maybe embarassing:
\begin{quotation}
\em 
``VaR is the \underline{minimum} potential loss that a portfolio can suffer in the 5\% worst cases in 7 days''
\end{quotation}
and is then frequently replaced by the more politically ``correct''
\begin{quotation}
\em 
``VaR is the maximum potential loss that a portfolio can suffer in the  \underline{95\% best} cases in 7 days''
\end{quotation}
A careful glance at the above definitions sheds some doubt on the fact that VaR is really the most appropriate variable to describe the risks associated with the 5\% worst cases of a distribution. In fact, once we have selected these cases, {\em why should we be interested in the least loss irrespectively of how serious all the other losses are ?} VaR, in other words, is a sort of ``best of worst cases scenario''  and it  therefore   systematically underestimates  the potential losses associated with the specified level of probability.

One could conclude that among the statistics  of   the 5\% left tail
VaR is maybe the less appropriate to characterize its risks.

Another paradoxical example that follows immediately from the above discussion is the fact that ``protection seller''--type portfolios  --- namely portfolios that earn small amounts with a high level of probability and suffer very large amounts with very small probability --- have negative 5\% VaR whenever the event they insure has a probability smaller than 5\%. VaR for such portfolios is completely unable to reveal any risk.

After this discussion and remembering the paradox of Section \ref{paradox} it seems vary natural to resort to some central estimate of the left 5\% tail rather than its maximum in order to construct a better variable to describe these risks. Minimizing imagination we therefore consider the possibility of introducing the {\em mean  of the 5\% worst cases}. 
We then define the Expected Shortfall as follows:
\begin{quotation}
\em 
``ES is the expected value of the loss of the portfolio in the 5\% worst cases in 7 days''
\end{quotation}
For the precise mathematical definition which requires some care, we refer to appendix A.

This variable has recently been studied and considered by some authors \cite{Embrechts,embrechts2}. Very surprisingly it was even considered in finance even before the VaR's age \cite{JPMorgan}.

It's clear that in the example of Section \ref{paradox}, the expected shortfall would perfectly distinguish between the levels of riskiness of portfolios A and B yielding for the latter a bigger value as anybody would expect from intuitive considerations.

In a comparison between VaR and ES for financial risk management purposes, the first crucial observation to make  comes from the above discussion: {\em ES is aware of the shape of the conditional distribution of 5\% worst events while VaR is not at all}.

In the following Section we are going to investigate a more technical and somehow unexpected difference between VaR and ES.

\section{Coherent Measures of Risk} \label{coherent}

A paper by  P. Artzner et al. \cite{artzner} face the problem  of defining a complete set of axioms that have to be fulfilled by a measure of risks in a generalized sense. A measure which satisfies these axioms is defined a 
``Coherent Measure of Risk''. It is then shown that whenever a portfolio is undoubtedly riskier than another one, it will always have an higher value of risk if the measure is coherent. On the contrary, any measure which does not satisfy some of the axioms will produce paradoxical results of some kind giving a wrong assessment of relative risks. 
In the paper, the class of coherent measures is identified and characterized and a coherence test is given. In the second axiom for a measure $\rho(\cdot)$,  we find a familiar concept:
\begin{enumerate}
\item[2] (Subadditivity) for any two random variables $X$ and $Y$:  $\rho(X+Y) \leq \rho(X)+\rho(Y)$.
\end{enumerate}
We already mentioned that VaR is subadditive in a Gaussian world and that this very property is one of the main themes of VaR supporters. Subadditivity in fact embodies in mathematical terms the reduction of risks associated with the concept of diversification. In most Risk Management textbooks some Gaussian example of VaR is mentioned where subadditivity is shown in such a way that the reader is led to believe that its validity in general. The problem is that it is not true despite this is today a firm belief among many finance practitioners.
{\em  VaR is not subadditive} apart from the Gaussian and some other special cases
 \cite{artzner}. Moreover, a little thought is enough to understand that in the Gaussian world everything is proportional to the standard deviation which in turn is subadditive. Therefore in the Gaussian world anything is subadditive and there's nothing special with VaR.

It's very easy to create an example of violation of VaR subadditivity: consider for example two different bonds $A$ and $B$ with non--overlapping default probabilities (think of two bonds issued by Nokia and Motorola: if one defaults the other will not and vice versa). A portfolio that contains both bonds may have a global VaR which is bigger than the sum of the two VaR's.
Consider for instance the following numerical example. The two bonds have two different  default states each with recovery values at 70 and 90 and probabilities 3\% and 2\% respectively. Otherwise they will redeem at 100.
\begin{center}
\begin{tabular}{|c|rrrc|}\hline 
\multicolumn{1}{|c}{\bf{final event}}			& \multicolumn{1}{|c}{\bf{A}}	&\multicolumn{1}{c}{\bf{B}}	& \multicolumn{1}{c}{\bf{A+B}} & \multicolumn{1}{c|}{\bf{Prob}}	 \\ \hline 
1	&70	&100	&170	&3\% \\
2	&90	&100	&190	&2\%\\
3	&100	&70	&170	&3\%\\
4	&100	&90	&190	&2\%\\
5	&100	&100	&200	&90\% \\\hline 
\end{tabular}
\end{center}
From these numbers it is easy to compute VaR and ES. For simplicity we suppose that the initial value of the bond is the expected value of the payoff in the above probability measure. VaR violation of subadditivity is shown below:
\begin{center}
\begin{tabular}{|l|rrr|}\hline
			& \multicolumn{1}{|c}{\bf{A}}	&\multicolumn{1}{c}{\bf{B}}	& \multicolumn{1}{c|}{\bf{A+B}}	 \\ \hline 
initial value	&98.9		&98.9		&197.8 \\
VaR 5\%   		&8.9		&8.9		&27.8\\
ES 5\%		&20.9		&20.9		&27.8\\
 \hline 
\end{tabular}
\end{center}
Notice that VaR fails because it underestimates the risks in the portfolios $A$ and $B$. To understand how serious this problem really is, we can imagine to minimize VaR for portfolio optimization purposes. The optimizator would produce as a result a portfolio fully invested in pure $A$ or in pure $B$ hence discouraging at all diversification\footnote{There's a close connection between subadditivity and convexity of the solutions surface in a problem of risk minimization.}!
Notice on the contrary that  ES in this example gives subadditive  results.

After the third paradoxical example on VaR results, we understand that we could go on forever inventing new examples of this kind, because the lack of coherence of this variable is proved in the example above where an axiom fails. 
 
But let's consider the impact that the lack of subadditivity can have on the use of VaR as a measure of firmwide Risk Management. We already mentioned the fact that computing firmwide VaR is often such a formidable task to perform that the reliability of the final results may be critical. The alternative which is in fact often used is to segment the computations by instrument and risk driver and to compute separate VaR's on branches and desks of a company. This is also sometimes necessary in financial institutions since the technological trading platforms are often distinct desk by desk. 
The problem is: once I know the separate VaR's of different branches, how can I make an estimate of global risks if VaR is not subadditive ? From the above example one can see that in principle global VaR may be much larger than the sum of partial VaR's. Nevertheless, the sum of different VaR's is commonly used as a best practice when global risks has to be assessed and global VaR engines are not available.

We show in Appendix A that our definition of Expected Shortfall gives a variable which satisfies subadditivity in full generality, with no assumptions at all on the probability distribution. 

The fact that in the examples above  ES produced sensitive results is in fact more than a coincidence. From the general definition of coherent measure we can in fact say in full generality that {\bf  it's impossible to built examples for which the assessment of relative riskiness among portfolios is trivial and in which at the same time  ES gives opposite results }.

\section{Conclusions}

This work starts from simple arguments that could be raised by every person who has worked with VaR. We have shown that no mathematics is needed to notice some evident flaws and to invent some variable more representative of the risks of extreme events of a portfolio distribution. Expected Shortfall appears as a natural choice to resort to when VaR is unable to distinguish between portfolios with different riskiness.

Only in a second moment the analysis becomes technical. With the concept of Coherent Measure of Risk 
we can compare the two variables and understand the source of all the problems of VaR paradoxes: its lack of subadditivity.

Finally we define a mathematically exact definition of Expected Shortfall and prove for this variable the subadditivity in full generality. Expected Shortfall is a coherent measure of risk while VaR is not as we show with a counterexample.

We think that ES is an excellent candidate for replacing VaR for financial Risk Management purposes, with many of the warnings that apply for VaR (delicacy of the estimate, reliability of approximations, transparency on hypotheses), but with the strong belief that, if correctly estimated, ES represents a solid instrument for assessing relative riskiness with no restrictions of applicability.

\newpage

\appendix

\section{Appendix: proof of subadditivity for the Expected Shortfall} \label{appendice1}

Let $F(x)$ be a probability distribution of a random variable $X$.
\begin{equation}
F(x) = Prob\{ X\leq x\}
\end{equation}
and for some probability $q \in (0,1)$ let's define the $q$-quantile
\begin{equation}
x_q = \inf \{x | F(x) \geq q\}
\end{equation}
If  $F(\cdot)$ is continuous we have $F(x_q)= q$, while if  $F(\cdot)$ is discontinuous  in $x_q$ and therefore $Prob\{X=x_q\}>0$ we may have  $F(x_q)= Prob\{X\leq x_q\} >q$. 
Our definition of $q$-Tail Mean $\overline{x}_q$ as the ``expected value of the distribution in the  $q$-quantile'' must take this fact into account.

\vspace{0.5cm}{\bf Definition:} {\em For a random  variable  $X$ and for a specified level  of probability $q$, let's define the  $q$-Tail Mean:}
\begin{equation}
\begin{array}{rcl}
\overline{x}_q &\equiv& \displaystyle \frac{1}{q} {\bf E}\left\{ X \;{\bf 1}_{X\leq x_q}\right\}
+ \left( 1 - \frac{F(x_q)}{q}\right) \, x_q \\
 &=& \displaystyle 
\frac{1}{q} {\bf E}\left\{ X \;{\bf 1}^q_{X\leq x_q}\right\}
\end{array}
\end{equation}
where in the last expression we introduced
\begin{equation}
{\bf 1}^q_{X\leq x_q} = {\bf 1}_{X\leq x_q} + \frac{(q-F(x_q))}{Prob\{X=x_q\}} 
{\bf 1}^q_{X = x_q}
\end{equation}
The second term in the sum is zero if $Prob\{X=x_q\}=0$.
In what follows we will make use of the following properties:
\begin{equation}  \label{prop1}
{\bf E}\left\{{\bf 1}^q_{X\leq x_q}\right\} = q 
\end{equation}
\begin{equation} \label{prop2}
0\leq {\bf 1}^q_{X\leq x_q}\leq 1 
\end{equation}
The only thing to show is (\ref{prop2}) in the case $X=x_q$:
\begin{equation}
\left. {\bf 1}^q_{X\leq x_q} \right|_{X=x_q} = 1 + \frac{q-F(x_q)}{Prob\{X=x_q\}} = \frac{q-F(x_q^-)}{Prob\{X=x_q\}} \in [0,1]
\end{equation}
since $F(x_q^-)\leq q \leq F(x_q^+) = F(x_q)$ and $Prob\{X=x_q\} = F(x_q)-F(x_q^-)$ by definition of $x_q$.

We may now prove the fundamental

\vspace{0.5cm}{\bf Theorem:} {\em given two random variables $X$ and $Y$ and defining $Z=X+Y$, one has:}
\begin{equation}
\overline{z}_q \geq \overline{x}_q +
\overline{y}_q  
\end{equation}
{\bf Dim:} 
\begin{equation}
\begin{array}{rcl}
q (\overline{z}_q - \overline{x}_q - \overline{y}_q ) &=& \displaystyle
{\bf E}\left\{
Z\; {\bf 1}^q_{Z\leq z_q} - X\; {\bf 1}^q_{X\leq x_q} -Y\; {\bf 1}^q_{Y\leq y_q}
\right\} \\\\
 & = & \displaystyle
{\bf E}\left\{
X\;\left(  {\bf 1}^q_{Z\leq z_q} -  {\bf 1}^q_{X\leq x_q} \right)
+
Y\;\left(  {\bf 1}^q_{Z\leq z_q} -  {\bf 1}^q_{Y\leq y_q} \right)
\right\} \\\\
&\geq & 
x_q \; {\bf E}\left\{
\left(  {\bf 1}^q_{Z\leq z_q} -  {\bf 1}^q_{X\leq x_q} \right)\right\}
+
y_q \; {\bf E}\left\{
\left(  {\bf 1}^q_{Z\leq z_q} -  {\bf 1}^q_{Y\leq y_q} \right)\right\}  \\\\
&=& x_q \, (q-q) + y_q\, (q-q) =0
\end{array}
\end{equation}
In the inequality we used the fact that
\begin{equation}
\left\{
\begin{array}{lll}
\displaystyle {\bf 1}^q_{Z\leq z_q} -  {\bf 1}^q_{X\leq x_q} \geq 0 & \hspace{5mm}\mbox{if} \hspace{5mm} & X>x_q \\\\
\displaystyle {\bf 1}^q_{Z\leq z_q} -  {\bf 1}^q_{X\leq x_q} \leq 0 & \hspace{5mm}\mbox{if} \hspace{5mm} & X<x_q 
\end{array}
\right.
\end{equation}
Which in turn is a consequence of (\ref{prop2}).

\noindent {\bf  CVD}

\vskip.1in

As a consequence we have: 

\vspace{0.5cm}{\bf Corollary:} {\em for any risk measure $R$ defined as}
\begin{equation}
R(X) = f(X) - \overline{x}_q
\end{equation}
{\em where $f$ is a linear function, the subadditive property holds}
\begin{equation}
R(X+Y) \leq  R(X)+R(Y)
\end{equation}

\vspace{0.5cm}{\bf Definition:} {\em For a specified time horizon $T$ let's define the $q,T$-Expected Shortfall $ES_{q,T}$, 
as the risk measure at time $t=0$ of a financial portfolio of value  $\Pi(t)$ defined by the difference between the forward price of the portfolio and the $q$-Tail Mean of the portfolio at time $T$.}
\begin{equation}
ES_{q,T}(\Pi) = 
D^{-1}(0,T)\, \Pi (0)
 - \overline{\Pi}_q(T)
\end{equation}

The above Corollary shows that the $q,T$-Expected Shortfall is subadditive.

\end{document}